_1

# Anisotropic Van der Waals 2D GeAs Integrated on Silicon Four-Waveguide Crossing

Ghada Dushaq*, Juan Esteban Villegas, Bruna Paredes, Srinivasa Reddy Tamalampudi, Mahmoud S. Rasras*

*Abstract*— **In-plane optical anisotropy plays a critical role in manipulating light in a wide range of planner photonic devices. In this study, the strong anisotropy of multilayer 2D GeAs is leveraged and utilized to validate the technical feasibility of on-chip light management. A 2D GeAs is stamped into an ultra-compact silicon waveguide four-way crossing optimized for operation in the O-optical band. The measured optical transmission spectra indicated a remarkable discrepancy between the in-plane crystal optical axes with an attenuation ratio of ~ 3.5 (at 1330 nm). Additionally, the effect of GeAs crystal orientation on the electro-optic transmission performance is demonstrated on a straight waveguide. A notable 50 % reduction in responsivity was recorded for devices constructed with cross direction compared to devices with a crystal *a*-direction parallel to the light polarization. This extraordinary optical anisotropy, combined with a high refractive index ~ 4 of 2D GeAs, opens possibilities for efficient on-chip light manipulation in photonic devices.**

## I. Introduction

RESEARCH in integrated photonics has been rapidly expanding to support advances in next-generation information technology requirements[1], [2]. Silicon photonics (SiPh) offers a cost-effective solution for a wide range of emerging applications in optical imaging[3], data communication[4], sensing, and quantum computing[5], [6]. The low cost and high integration density of current complementary metal-oxide semiconductor (CMOS) technology are very advantageous to SiPh integrated circuits. However, due to the physical limitations of silicon, it has not been an ideal photonic material for all kinds of on-chip devices including light sources and infrared detectors[7]–[9]. Despite the record-breaking demonstration of active silicon photonic components based on III-V hybrid integrated compounds and germanium[10]–[12]; the fabrication process still requires fabrication techniques to overcome the lattice mismatch with the silicon substrate[13], [14].

Over the past decades, two-dimensional (2D) materials have come to the forefront as viable building blocks of integrated photonic devices[15]–[17]. These materials have unique and fascinating optoelectronic characteristics. Furthermore, due to their multilayered nature, they are covalently bound in-plane and stack together out-of-plane via weak van der Waals force. As a result, the fabrication process can be greatly simplified and they can be integrated into a variety of substrates without being constrained by lattice matching[18], [19].

Recent advances in the heterogeneous integration of layered materials on silicon photonics (SiPh) platforms have been impressive[20]–[27]. For instance, graphene demonstrated an extraordinary performance for on-chip modulators and photodetectors[28], [29], while it has a universal absorption of ~ 2.3% in the visible and near-infrared spectrum[30]. Considering graphene is only 0.34 nm thick; this optical absorption efficiency is remarkable. Nevertheless, it is still too low to be effective for active photonic devices, which require higher absorption levels for effective functioning. Other reports on transition metal dichalcogenides (TMDCs) and their hetero structure were utilized for active photonic components that operate in the short-wavelength infrared (SWIR) ranges[21], [31]. However, most of their bandgaps fall inside the silicon absorption band which limit their integration benefits into SiPh, in addition to challenges in their hetero structure fabrication alignment[32], [33]. Black phosphorous (BP) and black arsenic phosphorus (BAsP) demonstrated a notable response in near and mid-infrared (NIR-MIR) spectrum [20], [23], [34], yet, BP is not chemically stable[35], [36]. Therefore, the exploration of innovative 2D materials with strong optical responses in the NIR-MIR, and excellent air stability is critical for the development of diverse integrated photonic devices.

The recently rediscovered germanium-based 2D materials such as GeP, and GeAs, distinguished themselves from the other candidates with their superior optoelectronic features[37]–[40]. For instance, they possess broadband detection due to the narrow and wide tunable bandgap energies (0.51–1.68 eV)[38], [41], [42], their integration capabilities on various substrate materials, and they exhibit sensitivity to the light polarization. According to our experiment and other theoretical research, these materials are chemically and dynamically stable[27], [41],

This work was supported by NYUAD Research Enhancement Fund (Corresponding authors: Ghada Dushaq & Mahmoud Rasras) Ghada Dushaq, Bruna Paredes, Srinivasa Reddy Tamalampudi and Mahmoud Rasras are with the Department of Electrical Engineering, New York University Abu Dhabi, UAE (e-mail: ghd1@nyu.edu ; bp64@nyu.edu; st4212@nyu.edu; mr5098@nyu.edu). Juan Esteban Villegas is with the Department of Electrical and Computer Engineering, New York University Tandon School of Engineering, NY, 11201, Brooklyn, USA (e-mail: jev1@nyu.edu ).

[42]. Furthermore, they have a monoclinic structure with low in-plane symmetry, which results in high electrical, optical[43]–[45] and thermal anisotropic properties[46]. Despite the outstanding optoelectronic and anisotropic characteristics of 2D GeAs, the majority of research is focused on theoretical studies and few demonstrations of surface illuminated and polarization-sensitive photodetectors in the visible range[47].

*In this study*, we demonstrate the first hybrid integration of anisotropic multilayer 2D GeAs on the SiPh platform for on-chip light manipulation. For this purpose, a dry transfer method was used to stamp 2D GeAs on an ultra-compact four-waveguide crossing (eight ports) that operates in the O-optical band. The paper is organized as follows, section II is devoted to the four-waveguide crossing design and fabrication, section III discusses the 2D GeAs integration and its optical properties, and section IV provides the experimental framework used to evaluate the performance of the integrated GeAs on silicon four-way crossing.

## II. WAVEGUIDE DESIGN AND FABRICATION

A waveguide four-way crossing structure based on silicon on insulator (SOI) platform with a 220 nm device layer and a 2 µm bottom oxide (BOX) is used. The whole structure is air-exposed (un-cladded). A false-color scanning electron microscope (SEM) image of the fabricated structure is shown in Fig. 1a. (zoom in SEM images are included in the supplementary material Fig. S1). The crossing is designed using topology optimization between the wavelengths from 1300 nm to 1800 nm. The design is optimized to have a small footprint using the open source lumopt adjoint optimization wrapper [48] and Lumerical finite-domain time-difference (FDTD) commercial software. The width of each crossing port is determined by a spline of ten symmetric points along the direction of propagation, in such a way that the position is modified in each iteration. Our design objective is to produce a waveguide crossing with high power transmission of the fundamental quasi-TE mode at the communication O-band. It is optimized with a maximum imbalance figure-of-merit of 0.639 (-1.95 dB).

**Fig. 1.** (a) false-color scanning electron microscopy image (SEM) showing the fabricated width of the tapered shape waveguide (b) simulated electric field intensity distribution of the optimized crossing for a quasi-TE mode polarization at 1310 nm (c) electric field intensity profiles ($|E|^2$) at the cut lines presented as points (1) to (6) in (a) to show the profile of the excited modes in the crossing regime (d) the calculated transmission and reflection spectra at 45˚, 90˚, and 135˚ outputs of the optimized design.

Similar to other waveguide crossings, the design pushes the imaging condition close to its geometric center in the optimized state[49]. However, the small footprint (constrained at 6.5 x 6.5 µm) forces the coupling into higher-order optical modes to have good interference while crossing. This results in relatively high scattering losses, yet very low coupling crosstalk into other ports.

Figure 1b shows the electric propagating field intensity profile of the fundamental transverse electric ($TE_{00}$) mode in the waveguide crossing. Additionally, the normalized intensity profiles ($|E|^2$) at various waveguide point locations from (1) to (6) in Fig.1a are shown in Fig.1c. As can be seen, the electric field experiences a multi-interference at the crossing region, and afterward it passes to the output waveguide. It is worth mentioning that the fraction of TE polarization at the output waveguide is 0.99 i.e. ~99.2% of the light is still TE polarized. The simulated performance of our waveguide crossing is depicted in Fig.1d. An insertion loss and return power of 2.50 dB and -21.81 dB were recorded, respectively. Additionally, the coupling into each of the ports at 45°, 90°, and 135° (crosstalk) from the input is -23.57 dB, -24.75 dB, and -24.92 dB, respectively. The measured insertion loss and crosstalk spectra of the structure are presented in the supplementary material Fig. S1. A crosstalk and insertion losses of -22.8 dB and 2.7 dB were measured, respectively. These values are in good agreement with the simulation presented above. More details on the optical transmission with and without material integration in the wavelength range of 1290 nm–1330 will be discussed in section IV.

## III. GeAs INTEGRATION AND OPTICAL PROPERTIES

### A. GeAs Anisotropy

Bulk GeAs is a monoclinic crystal with space group C2/m where the structure is formed in such a way that the adjacent layers are stacked along the c-direction by week van der Waals forces[41], [43], [50]. The in-plane armchair and zigzag directions are associated with the crystal b-axis and *a*-axis, respectively (crystal structure is described in the supplementary information Fig.S2). Typically, 2D GeAs exhibits a strong optical and electrical anisotropy due to their low structure symmetry[43]. Hence, light absorption and carrier mobility in GeAs vary with the crystal orientation.

The multilayer GeAs flakes were exfoliated from their bulk crystals and transferred on top of the four-waveguide crossing structure by using a deterministic dry transfer process [21], [51], [27], [52]. The flakes were commercially available from 2D Semiconductors. In this study, we transferred GeAs flakes with thicknesses in the range of 32 nm to 80 nm. Figure 2a shows the atomic force microscopy scan of the transferred flake with a thickness of ~ 75 nm. The scan depicts the conformal coverage and strong adhesion of the multilayer GeAs on the top of the four-waveguide crossing.

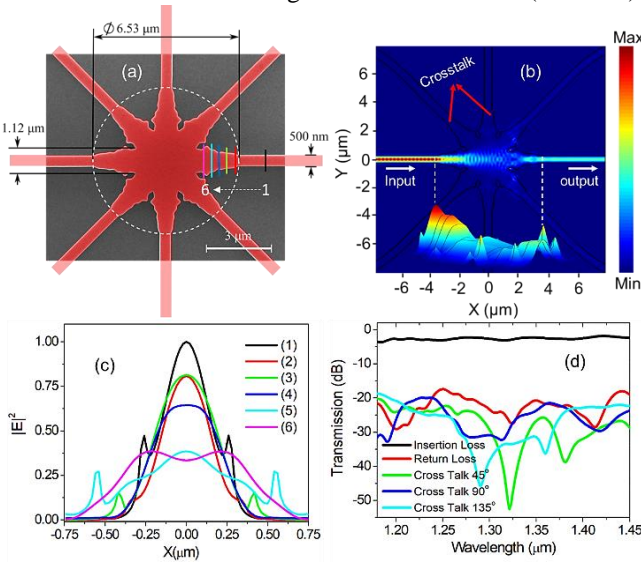

*B. Angle-Resolved Polarized Raman Spectra (ARPRS)*

To label the in-plane *a*-and b-axes of the transferred flake, angle-resolved polarized Raman spectra (ARPRS) measurements have been carried out in the parallel and perpendicular configurations as shown in Fig. 2b. The Raman spectra was recorded using a WITECalpha500 Confocal Raman system in an ambient air environment. It is worth noting that the laser polarization was fixed, and the analyzer direction was aligned either parallel or perpendicular to the laser polarization. The zero position of the sample, i.e. the angle between the laser polarization and crystal direction is equal to zero, was set across the $I_4O_1$ direction (*b*-direction, see Fig.2a), then the ARPRS measurements proceeded by rotating the sample from 0° to 360° with 10° steps.

The polar plots of Raman peak intensity are recorded for the two out-of-plane phonon modes of 147 cm$^{-1}$ and 174 cm$^{-1}$ under parallel and perpendicular configurations (Raman spectra in the parallel and cross direction are included in supplementary material Fig. S2). As can be seen, the Raman peak intensity shows a strong sensitivity to the crystal orientation, polarization angle, and analyzer configuration. The maximum and minimum Raman intensity of the 147 cm$^{-1}$ phonon mode was observed at 90° and 0° in the parallel configuration, respectively. For the same phonon mode, the maximum and minimum intensity were changed to 150° and 40° in the cross configuration without any change in the peak position. The above results are in good agreement with previous reports which show a similar dependency of the 147 cm$^{-1}$ and 174 cm$^{-1}$ phonon vibrational modes on the polarization angle[43].

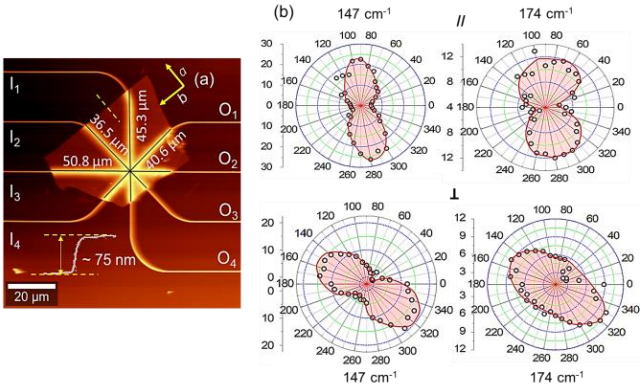

**Fig. 2.** 2D GeAs integration and characterization (a) Atomic Force Microscopy (AFM) image scan of the transferred multilayer GeAs on the four-waveguide crossing, the dashed yellow line shows a thickness of ~ 75 nm GeAs flake (b) polar plots of angle-resolved polarized Raman spectroscopy (ARPRS) measurements of the fitted peak intensities of $A_g$ modes at 147 cm$^{-1}$ and 174 cm$^{-1}$ performed on the flake scanned in (a), Black circles are the experimental data and the red solid line is an eye guideline.

*C. GeAs Optical Properties*

The optical parameters of 2D GeAs are extracted from high-resolution Accurion ellipsometry measurements and plotted in Fig. 3a. Details on experiment and modeling with different flake thickness are included in the supplementary material Fig. S3 and Table S1.

The extracted $n$ and $k$ values, (real and complex index of refraction, respectively) of the multilayer GeAs were fed into the Lumerical mode solver software. Figure. 3b shows the simulated electric-field profile ($|E|^2$) of $TE_{00}$ mode for the 65 nm GeAs/Silicon waveguide heterostructure at 1310 nm wavelength. The optical mode interaction for other flake thickness is included in the supplementary material Table S2. It is noted that the $TE_{00}$ optical mode of the Si waveguide overlaps with the multilayer GeAs flake. The 65 nm thick flake's optical transmission loss is estimated to be 2.6 dB/µm at 1310 nm.

The optical losses calculated for other flake thickness are included in the supplementary material. Additionally, we performed a Mode Expansion Beam PROP simulation (EME) to study the attenuation as a function of the flake's interaction length (see Fig. 3b). It is noted that a significant attenuation (near-complete absorption) for the 65 nm thick flake with an interaction length <10 µm at 1310nm wavelength. These results are consistent with our previous reports on SWIR GeAs photodiode [41]. It is also worth mentioning that integrated GeAs on the Si waveguide offers enhanced light-matter interaction compared to the free space coupling.

More details on the Beam Prop simulation for different GeAs layer coverages and thicknesses can be found in the supplementary material Fig. S4 and Fig. S5.

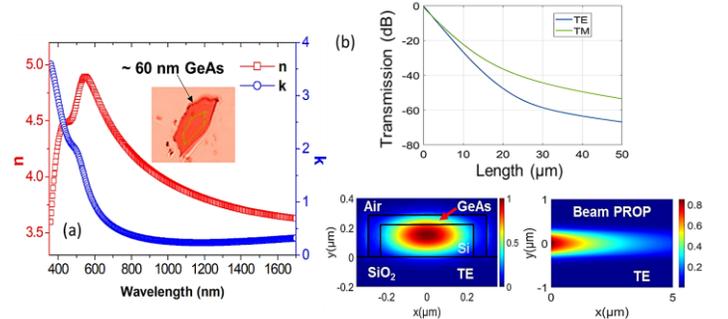

**Fig. 3.** (a) The measured real and complex part of the optical index of refraction for ~ 60 nm GeAs flake recorded using high-resolution ellipsometer (b) Electric-field intensity profile ($|E|^2$) of 65 nm GeAs layer on Si at 1310 nm and the optical losses of the same flake thickness as a function of interaction length.

## IV. MEASURED WAVEGUIDE CROSSING RESPONSE

*A. Passive Optical Response*

The optical transmission of the four-waveguide crossing was initially characterized without GeAs flakes to determine its intrinsic losses. A transverse electric (TE) polarized light is edge coupled into the crossing through a lensed fiber using a tunable laser operating at O-bands, while the output response is collected by an output lensed fiber and detected by a power meter. The measured total insertion loss of the input ($I_i$) and output ($O_j$) in the $I_1O_4$, $I_2O_3$, $I_3O_2$, $I_4O_1$ (See Fig. 2a) ports were in the range of 20 - 21 dB at a wavelength of 1290 nm -1330 nm.

Next, multilayer GeAs flakes with a thickness range of ~ 46-76 nm were transferred on the nanowire crossing structures (see Fig. 4a). Figure 4b shows an optical image of ~ 46 nm GeAs flake transferred on the nanowires structure. Light in the Si waveguides is evanescently coupled to the GeAs flake. The excess optical losses for the TE polarization induced by flakes were determined by subtracting the loss of the reference crossing from that obtained in the hybrid structure (GeAs/Si). Here, the supported TE mode in the Si waveguide has an electric field along the x-direction (perpendicular to the light propagation direction). Herein, the optical losses are mainly associated with material absorption. Losses due to reflection and scattering (mode mismatching losses) at the coupling interface between the passive waveguide (air/Si) to the active region (GeAs/Si) have a negligible effect as presented in the supplementary material Fig. S6.

ARPRS measurements were carried out on the transferred flake as shown in Fig. 4c. The light attenuation is highly affected by the crystal orientation, where the strongest absorption occurs when the light polarization is along the *a*-direction of GeAs crystal while it is weakest when the polarization is parallel to the b-direction.

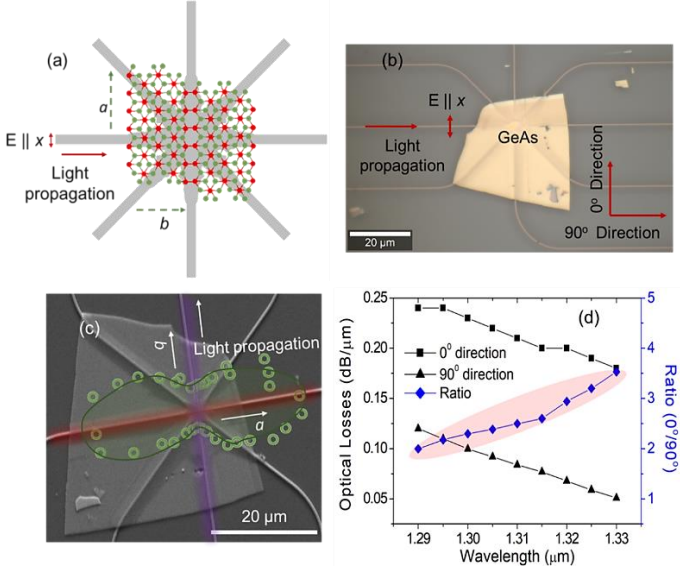

**Fig. 4.** Crystal orientation and light attenuation of 2D GeAs (a) schematic of 2D GeAs crystal structure on the four nanowire crossing (b) optical image of ~ 46 nm GeAs on the crossing showing light propagation and polarization directions (c) SEM image of the GeAs showing the Raman intensity variation with the crystal orientation (d) the measured normalized optical losses at O-optical band in the 0˚ and 90˚ direction of the crystal as depicted in (b), additionally, the second y-axis shows the ratio obtained between them.

The normalized optical losses in the *a* (0°) and *b* (90°) directions (see Fig. 4c) are plotted in Fig. 4d. The measured optical losses at 45° and 135° and for other device thicknesses are listed in supplementary material Tables S3 and S4. As can be noticed, the optical losses decrease with a wavelength which is consistent with the material absorption trend observed in our previous study[41]. On the other hand, the *a*-direction exhibits a significant optical attenuation compared to the b-direction with a transmission ratio of 3.5 at 1330 nm. This behavior is attributed to the distinct anisotropic band dispersion of the GeAs flakes. This results in variation of optical transition rates at different crystal directions, leading to the observed direction-dependent optical absorption. Based on these results, the 2D GeAs acts as a density optical filter in such a way that the intensity transmission in the O-band can be effectively tuned by modulating the crystal orientation. This can be of high interest in high-density integrated photonics circuit applications when one often requires to attenuate intense light before it hits a certain photodetector, or to attenuate a laser beam. Additionally, taking advantage of this remarkable light attenuation and the strong in-plane optical anisotropy of multilayer GeAs, we believe it can offer a new venue for multi-functional on-chip optical filters and switches.

*B. Active Optical Response (Photodetection)*

The observed significant variation in the GeAs absorption with crystal orientation and polarization direction highly affects the performance of the photoelectric response. Therefore, these effects are tested at 1310 nm light. In the experiments, we adopted a straight waveguide with a width of 460 nm. Two metal pads of Ti/Au (10 nm/120 nm) were symmetrically deposited around the waveguide as electrodes, see Fig. 5a. The source-drain electrodes are designed with a 3–5 µm gap at both sides of the silicon waveguide. Additionally, two ~50 nm flakes with different crystal orientations were transferred on top of the straight waveguide and laid directly on the source-drain electrodes as shown in Fig. 5a and b. Device (1) has its crystal orientation and light propagation in the waveguide along the zigzag direction (*a*) of GeAs while device 2 is perpendicular.

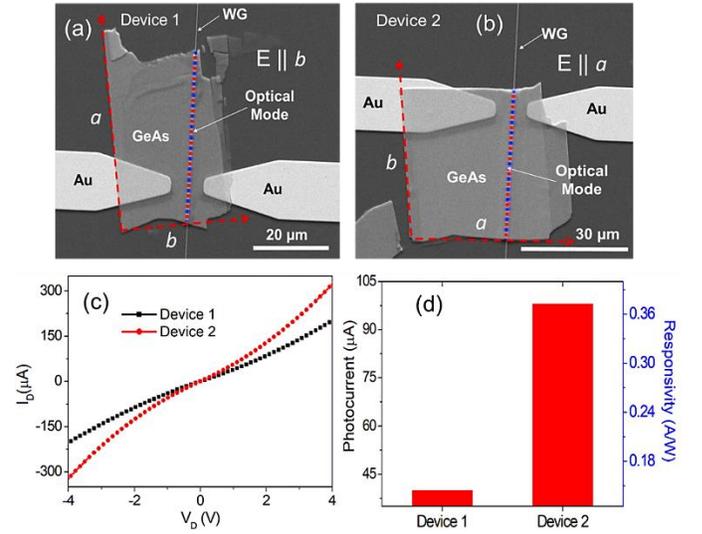

**Fig. 5.** Steady-state photoelectric response (a)-(b) SEM images of integrated multilayer GeAs (~ 50 nm) on a straight waveguide with two different crystal orientation alignments with light propagation (c) I-V characteristics under dark conditions for the two devices (d) Photocurrent and responsivity response recorded for the two devices when 1310 nm TE polarized light is edge coupled into the devices.

Figure 5c shows the I-V characteristics of the two devices, where device (1) exhibited a lower dark current (conductance) compared to device (2). This result is in agreement with other studies on anisotropic electronic transport where higher mobility and conductance along the zigzag direction of GeAs crystals are observed[43]. Figure 5d, depicts the devices responsivity when a TE polarized light is edged coupled into the waveguide using lensed fiber in such a way that light propagation is parallel to the crystal *a*-direction in device (1) and to the crystal b-direction in device (2) (see Fig. 5a and b). Results show that device (2) achieved a higher responsivity of ~ 0.36 A/W. In this device the electric field and the carrier transport direction are along the *a*-direction of the flake and perpendicular to the light propagation direction. In contrast, device (1) exhibited a ~ 50% reduction in the responsivity where the electric field and the carrier collection direction are along the b-direction of the crystal. As a result, the total light attenuation is dependent on the orientation of the GeAs (active layer) with respect to the passive Si waveguide underneath. Additionally, it is noteworthy to point out that the devices exhibited a broadband operation spectral responses which demonstrates the wide response of GeAs crystals. A responsivity of ~ 0.4 A·W$^{-1}$ is observed when an input optical power of 200 µW is coupled to the device at 4V bias.

## V. Conclusion

In summary, the anisotropy of 2D GeAs flake is demonstrated using four-waveguide silicon photonics crossing structure. For this purpose, we designed and fabricated an ultra-compact and low loss four-waveguide crossing (eight ports) optimized for the fundamental quasi-TE mode operation. In the devices, a dry transfer method was utilized to stamp 2D GeAs into the waveguide crossing. Additionally, high resolution ellipsometry measurements are used to extract the optical parameters where it is observed that 2D GeAs exhibits a high index of refraction of ~ 4 and an estimated absorption coefficient of 2.6 dB/µm measured for 65 nm thick flake at 1310 nm.

Furthermore, the measured optical transmission spectra of the hybrid-integrated GeAs crossing demonstrated a remarkable light attenuation deference between the two orthogonal in-plane crystal axes orientations. During measurements, a maximum attenuation ratio difference of ~ 3.5 is observed. This unprecedented in-plane optical anisotropy was also tested by measuring the photo response performance of integrated GeAs devices. It is shown that a considerable photoresponsivity of 0.4 A/W can be obtained for devices with crystal *a*-direction parallel to the light polarization compared to a 50 % reduction in devices constructed with crystal b-direction parallel to the light polarization. We believe this integrated system can be employed in applications related to on-chip optical density filters, optical switches, artificial synaptic devices, and SWIR polarization sensitive photodetectors.

## Acknowledgment

The authors are thankful to NYUAD Photonics and Core Technology Platform Facility (CTP) for the analytical, material characterization, devices fabrication, and testing. The first author acknowledges L'Oréal UNESCO For Women in Science Middle East Fellowship. We are also grateful to Dr. P. H. Thiesen a senior application specialist at Accurion for helping us in testing the optical parameters of 2D GeAs.